\documentclass[10pt,a4paper]{article}
\usepackage[utf8]{inputenc}
\usepackage[ukrainian,english]{babel}
\usepackage{amsmath}
\usepackage[pdftex]{graphicx}
\usepackage{epstopdf}
\usepackage{caption}
\usepackage{subcaption}
\usepackage{times}
\usepackage{hyperref}
\usepackage{float}
\usepackage{indentfirst}
\usepackage{geometry}
\usepackage{array}        

\begin{document}

\begin{center}
\fontsize{12}{14}\selectfont
\noindent \textbf{DETERMINATION OF THE TOTAL DOSE OF BREMSSTRAHLUNG X-RAY REMINISCENCE ON THE HIGH-CURRENT PULSED RADIATION-BEAM COMPLEX "TEMP-B"}
\end{center}

\begin{center}
\fontsize{12}{14}\selectfont
\noindent \textbf{A.B. Batrakov, S.I. Fedotov, I.N. Onishchenko, E.G. Glushko, A.M. Gorban, O.L. Rak,}

\noindent \textbf{O.V. Nevara, Yu.N. Volkov}
\end{center}
\begin{center}
\fontsize{10}{12}\selectfont
\noindent {\emph{National Science Center "Kharkiv Institute of Physics and Technology", Kharkiv, Ukraine}}

\text{\emph{E-mail: a.batrakov67@gmail.com}}
\end{center}

The paper reports the results of measuring the total dose of X-ray bremsstrahlung from a powerful X-ray source based on the high-current pulsed direct-action electron accelerator “Temp-B”. The parameters of the high-current, tubular relativistic electron beam from the accelerator were as follows:  energy 600 keV, current 13.5 kA, and pulse duration 1.0$\mu$s. Using the pulsed magnetic field of a solenoid, the electron beam generated in a magnetically isolated diode was transported over a 55 cm distance toward the molybdenum converter. An auxiliary coil, connected in series with the solenoid, was placed adjacent to it to provide the desired magnetic field in the converter region and avoid beam losses. The methodology for determining the total dose of the produced X-ray bremsstrahlung is described. Polycrystal-line detectors were used for measuring the X-ray bremsstrahlung dose. They were located 80 mm behind the converter in the polar plane, with a $20^\circ$ separation. Measurements of the dose distribution over the polar angle showed symmetrical distributions of the radiation at both polar and azimuthal angles relative to the axis. Taking into account such symmetry and the fact that the electron beam radiates solely into the forward hemisphere, the integration over the entire sphere can be reduced to integration over just a fraction of 1/8 of the spherical surface, thereby reducing the required number of sensors. The experimentally obtained value of the total dose of the X-ray bremsstrahlung is 388.5 Gy per single pulse of the accelerator current, which is concentrated in a $120^\circ$ cone in the direction of the beam movement.

\begin{flushleft}
\fontsize{10}{12}\selectfont
\noindent \textbf{Keywords:} \emph{High-current pulsed direct-action electron accelerator; Relativistic electron beam; Converter; X-ray bremsstrahlung, Thermo-luminescent detector; Intensity, Dose.}

\noindent \textbf{PACS:}  41.50.+ h; 41.60.Gr
 \end{flushleft}

\begin{center}
\fontsize{12}{14}\selectfont
\noindent \textbf{INTRODUCTION}
\end{center}

Charged-particle accelerators and gamma-emitting isotopes are widely used to produce powerful fluxes of hard X-ray radiation. Isotopic sources are structurally simpler, while accelerator-based sources deliver higher radiation powers, which is essential for their integration into high-performance production lines. To generate bremsstrahlung X-rays with relativistic electron beams (REBs), two types of converters are used in the form of transmitting and reflecting targets, which expands the possibilities of controlling the radiation flux. The hard X-ray radiation generated by REBs is widely used in industrial processes, security systems, and non-destructive testing technologies due to its great penetration depth and high spatial resolution. In addition, there has been recent interest in the use of powerful pulsed X-ray sources in medicine [3]. Currently, extensive work is underway on both the creation of new X-ray sources based on direct-action electron accelerators [4 - 7] and the improvement of methods for transporting and controlling the X-ray flux [8]. A feature of most such installations is the placement of the converter directly at the output of the accelerating diode, or its manufacture as a diode element. However, sometimes it is desirable to place the converter remotely from the accelerator, which requires transporting the electron beam over a considerable distance. At NSC "KIPT", for quite some time, work has been carried out on the creation and operation of appropriate radiation-beam complexes [1, 9 - 11]. This work presents the design of a radiation-beam complex for the development of a powerful bremsstrahlung X-ray source based on the high-current pulsed electron accelerator "Temp-B", with the converter located remotely from the magnetically isolated accelerator diode, and the results of measuring the total radiation dose per pulse.

\begin{center}
\fontsize{12}{14}\selectfont
\noindent \textbf{EXPERIMENTAL SETUP}
\end{center}

The accelerator "Temp-B" consists of the following main elements: a pulsed voltage generator (PVG), a magnetically isolated vacuum diode, a chamber for REB transport, a converter, a magnetic system, and devices for recording accelerator parameters and X-ray bremsstrahlung radiation. The PVG provides a 600 kV pulse with a duration of 1.0$\mu$s. To reduce the inductance of the PVG and the current load on the discharge electrodes, a design consisting of 4 parallel Marx generators is implemented, placed in a metal tank filled with transformer oil. In this case, the current load on the spark gap is reduced by 4 times, and the total inductance of the PVG is 1.5 times smaller. To reduce the total number of spark gaps by 2, a bipolar charging system is used ($\pm$100 kV instead of a unipolar 100 kV system) with two capacitors in series per stage. This successfully solves the problem of simultaneously switching the generators onto a standard load.

The source of the high-current REB in the accelerator is a magnetically isolated vacuum diode with an explosive-emission cathode. The design of the vacuum diode with magnetic isolation is shown in Fig. 1.

\begin{figure}
  \centering
  \includegraphics[width=0.5\textwidth]{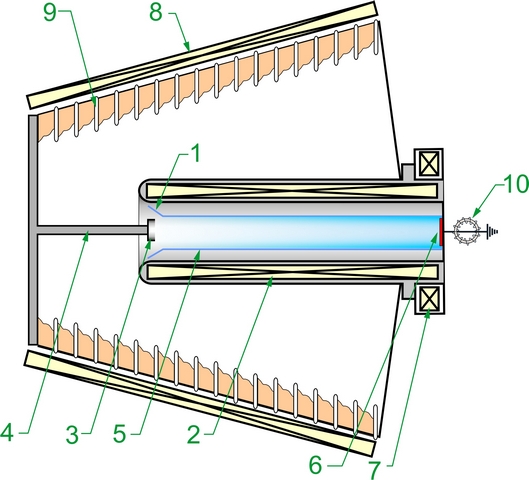}\\
  \caption{\emph{Scheme of the diode and the chamber for beam transport.}}\label{fig:image_label}
\end{figure}

It is assembled according to a scheme in which the anode (1) is a conical insert connected to the beam transport chamber (5). The use of an anode insert allows for electron flow parallel to the uniform external magnetic field in the drift chamber. The magnetic isolation of such a diode is carried out by the magnetic field of the left edge of the solenoid (2). The sharp front edge of the tubular cylindrical cathode (3), of a 60 mm diameter, mounted on the cathode holder (4), is located in the region of the increasing magnetic field of the solenoid (2).

The beam-transport chamber (5) is a thin-walled stainless steel cylinder with an inner diameter of 70 mm and a length of 550 mm. It connects the diode to the converter region (6). The chamber is placed inside the solenoid of the guiding magnetic field (2), which has 188 turns arranged in two layers, with a 150 mm diameter. For the current pulse of 2 kA and duration of 10 ms feeding the solenoid, the magnetic field strength in the beam transport chamber is 640 kA/m. Furthermore, an auxiliary coil (7), which has 34 turns in two layers at a diameter of 250 mm and is connected in series with solenoid, is located in the interaction chamber region. It is designed to provide the necessary magnetic field strength in the converter region. The calculated and experimentally implemented magnetic field parameters ensured magnetic insulation and the transport of the electron beam with a current of 13.5 kA and an energy of 600 keV to the converter without losses. Outside the vacuum diode chamber, a conical, two-layer insulating magnetic field coil (8) with 208 turns is located. It is powered by a separate source and is intended to increase the electrical strength of the accelerating column insulator (9). A Rogowski coil (10) was used to measure the beam current.

\begin{figure}[H]
  \centering
  \includegraphics[width=0.9\textwidth]{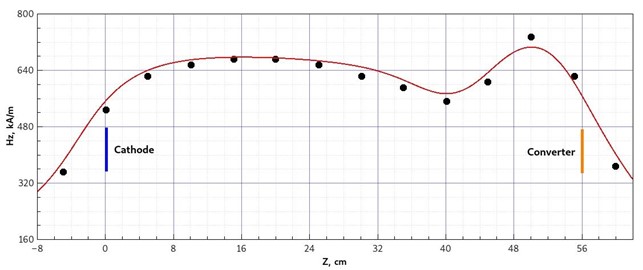}\\
  \caption{\emph{Profile of the longitudinal magnetic field in the chamber for beam transporting.}}\label{fig:image_label}
\end{figure}

Fig. 2 shows a graph of the calculated longitudinal magnetic field profile (solid curve) in the beam transport chamber. It shows the results of summing the fields from the solenoid, the auxiliary coil, and the conical insulating field coil. On the graph, the origin of coordinates is aligned with the cathode's end face, and the converter is located 56 cm from the cathode. Experimentally measured magnetic field strengths are marked with dots. To calculate the magnetic field strength of the solenoid and a combination of several coils, they are considered as a set of ring currents, and the magnetic field at a given point is determined by the total contribution of the magnetic fields of all ring currents, according to [12]. In this case, all currents are assumed to be stationary, and the possible influence of other structural elements is not considered. The calculations are consistent with the measurements within the apparatus's 5\% error.

\begin{center}
\fontsize{12}{14}\selectfont
\noindent \textbf{EXPERIMENTAL RESULTS}
\end{center}

In Fig. 3, the voltage applied to the magneto-isolated diode is shown. A 60 mm-diameter, one mm-thick tubular cathode was used to produce a relativistic electron beam. The beam, with a diameter of 53 mm, energy of 600 keV, current of 13.5 kA, and duration of 1 $\mu$s, was delivered from the diode by the magnetic field to a molybdenum converter of a 0.5mm wall thickness. Figure 4 shows a photograph of the beam imprint on the surface of the molybdenum converter after two electron beam pulses. The good repeatability of the imprints should be noted. At a converter plate thickness of 0.5 mm, signs of the onset of its destruction process are clearly visible.

\begin{figure}[H]
  \centering
  \begin{minipage}{0.45\textwidth}
  \includegraphics[width=0.95\linewidth]{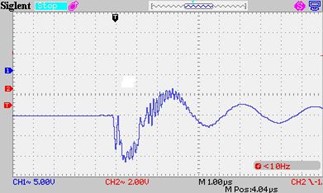}\\
  \caption{\emph{Oscillogram of the diode voltage.}}
  \label{fig:image_label}
\end{minipage}
\hfill
  \begin{minipage}{0.45\textwidth}
  \centering
  \includegraphics[width=0.6\textwidth]{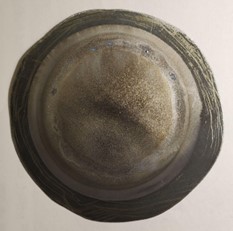}\\
  \caption{\emph{Electron beam imprint on the molybdenum converter surface.}}
  \label{fig:image_label}
\end{minipage}
\end{figure}

Thermo-luminescent detectors (TLDs) of types MTS-6 and MTS-7 [14] were used to determine the angular distribution of the X-ray bremsstrahlung radiation resulting from the interaction of the relativistic electron beam with the converter. These detectors are polycrystalline LiF "tablets", doped with Mg and Ti atoms with a dose measurement range from $10^{-4}$ Gy to 1 Gy in the X-ray and gamma radiation energy range, specifically from 20 keV to 6 MeV. Their diameter is 3.5 mm, area equals $9.62 \text{ mm}^2$, and thickness is 2 mm. Such detectors allow the measurement, with appropriate calibration, of absorbed doses over a wide energy interval. The measured results were processed using the automatic device RADOS TLD RE-2000 [13], which automatically reads the values stored in the dosimeters. The X-ray detectors were placed 8 cm from the converter and positioned at 20-degree intervals relative to the electron beam axis. A total of 5 detectors were used (Fig. 5).

\begin{figure}[h]
  \centering
  \includegraphics[width=0.5\textwidth]{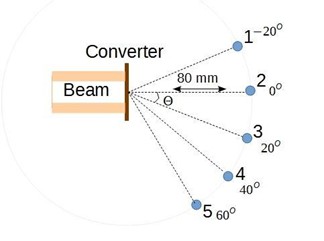}\\
  \caption{\emph{Disposition of the X-ray radiation detectors.}}\label{fig:image_label}
\end{figure}

Table 1 presents the results of measuring the generated dose absorbed by the detector in a series of two electron beam pulses. The detector numbers correspond to those indicated in Fig. 5.

\begin{table}[h]
\centering
\caption{Results of measuring the generated dose, absorbed by the detector.}
\begin{tabular}{|c|c|c|c|c|c|}
\hline
Detector Number & 1 & 2 & 3 & 4 & 5 \\
\hline
Angle relative to beam axis, degrees ($^\circ$)  & $-20^\circ$ & $0^\circ$ & $20^\circ$ & $40^\circ$ & $60^\circ$ \\
\hline
Absorbed Dose (from RE-2000 reader), mGy & 403,643 & 417,884 & 410,272 & 211,88 & 181,332 \\
\hline
\end{tabular}
\end{table}

The measurement results for detectors 1 and 3 are approximately the same, indicating a symmetrical distribution of X-ray radiation relative to the polar angle $\Theta$ in the azimuthal plane where the detectors are located. For angles greater than $60^\circ$, it was impossible to place the detectors due to structural reasons of the source. Nevertheless, a decrease in dose with increasing angle is observed, such that the radiation in the $120^\circ$ cone is close to the radiation dose in the forward hemisphere, and therefore to the total dose of the source, as the beam only radiates into the forward hemisphere.

To measure the total radiation dose in the forward hemisphere, it is sufficient to measure its distribution over a spherical surface of a given radius. The total X-ray dose was calculated as follows. Considering that the element of area dS of a sphere with radius r is:

\begin{center}
\fontsize{12}{14}\selectfont
\noindent \emph{dS=$r^2$sin$\Theta$d$\Theta$d$\varphi$}
\end{center}

\begin{flushleft}
where $\Theta$ and $\varphi$  are the polar and azimuthal angles, respectively, the dose value $D$, distributed over the hemisphere, is:
\end{flushleft}

\begin{equation}
    D = \int\limits_{0}^{2\pi} \,d\varphi \int\limits_{0}^{\frac{\pi}{2}} \,d(\Theta)r^{2}\sin\Theta d \Theta
\end{equation}

Here, d($\Theta$) is the surface dose density on the sphere of radius $r$. All detectors are placed in the polar plane. Since d($\Theta$) depends only on $\Theta$, and the integration over $\varphi$ simply gives a factor of $2\pi$, due to the azimuthal symmetry of the radiation distribution, Eq.(1) can be written as:

\begin{equation}
     D = 2{\pi} r^{2} \int\limits_{0}^{\frac{\pi}{2}} \,d(\Theta)\sin\Theta d \Theta
\end{equation}

The dose $d_{i}$, measured by a detector of an area $S=9.62$ $mm^2$ is a discrete estimate d($\Theta$)=$d_{i}/$S.

To calculate the total dose, we use the trapezoidal rule. In our case, the last detector in the row is placed at an angle of $60^\circ$. Therefore, we estimate the dose within the limits (in degrees) -60$\leq\Theta\leq$60, such that

\begin{equation}
    D {\simeq} {\frac{2{\pi} r^{2}}{S}} \cdot \frac{{\pi}/2} {3} \Biggl[\frac{f_{2}+f_{5}}{2}+f_{3}+f_{4}\Biggr]=2188,685\cdot\Biggl[\frac{f_{2}+f_{5}}{2}+f_{3}+f_{4}\Biggr]{\simeq}777,058 Gy
\end{equation}

The magnitudes assumed by members of the integrand in (2) are given in Table 2.

\begin{table}[h]
\centering
\caption{Initial data for calculating the radiation dose.}
\begin{tabular}{|c|c|c|c|}
\hline
Detector Number & Angle & Dose in detector $d_{i}$, mGy & $f_{i}=d_{i} \cdot \sin \Theta$ \\
\hline
2 & $0^\circ$ & 417,884 & 0 \\
\hline
3 & $20^\circ$ & 410,272 & 140,321 \\
\hline
4 & $40^\circ$ & 211,88 & 136,194 \\
\hline
5 & $60^\circ$ & 181,332 & 157,038 \\
\hline
\end{tabular}
\end{table}

Thus, the dose of X-ray radiation limited by the source measurement angle of $120^\circ$, which is close to the full dose of radiation from the source, per one electron beam pulse with an energy of 600 kV, a current of 13.5 kA and a duration of 1 $\mu$s, is D=777/2=388.5 Gray/pulse.

\begin{center}
\fontsize{12}{14}\selectfont
\noindent \textbf{CONCLUSION}
\end{center}

The doses of X-ray bremsstrahlung radiation, realizable per one current pulse from the high-current "Temp-B" accelerator, have been estimated for a variety of angles relative to the direction of the relativistic electron beam’s motion.

The magnitude of the total radiation dose has been estimated. For the beam of a 600 keV energy, a 13.5 kA current magnitude, and a 1 $\mu$s duration, the total radiation dose from the source was 388,5 Gy/pulse.

As found, most of the X-ray radiation is confined within a $120^\circ$ cone about the beam propagation direction.

\begin{center}
\fontsize{12}{14}\selectfont
\noindent \textbf{REFERENCES}
\end{center}

\begin{flushleft}
[1]	Yu.V. Tkach, V.T. Uvarov, N.P. Gadetskiy, et. al. Abstracts of reports. (NSC KIPT, Kharkiv, 1983). p.107, (in Russian)

[2]	T.C. Genoni, and R.B. Miller, “Limiting currents in shielded source configurations,” Physics of Fluids, 24(7) 1397–1398 (1981).

[3]	H.S.Li, R. Tang, H.S. Shi, et al.  Sig. Transduct. Target Ther. 10(82), (2025). https://doi.org/10.1038/s41392-025-02184-0

[4]	H. Shi, C. Zhang, P. Zhang, Y. Wang, W. Yuan, L. Chen, X. Li, J. Wu, and A. Qiu,  IEEE Trans. Plasma Sci. 51(4), 1142–1149 (2023). https://doi.org/10.1109/tps.2023.3254654

[5]	P. Zhang, H. Shi, Y. Wang, C. Zhang, M. Xu, D. Wang, X. Li, et al. J. Appl. Phys. 133, 243301 (2023). https://doi.org/10.1063/5.0151604

[6]	Lai Dingguo, Qiu Mengtong, Yang Shi, et al. High Power Laser and Particle Beams, 33, 035001 (2021) https://doi.org/10.11884/HPLPB202133.200269

[7]	M. Seltzer, J.P. Farrell, and J. Silverman.  IEEE Transactions on Nuclear Science, 30(2), 1629-1633 (1983) https://doi.org/10.1109/TNS.1983.4332602

[8]	M. Vassholz, and T. Salditt. Science Advances, 7(4), (2021) https://doi.org/10.1126/sciadv.abd5677

[9]	A.B. Batrakov, E.G. Glushko, A.M. Yegorov, et al.  Problems of Atomic Science and Technology. (6(100)), 100-104 (2015).

[10] A.A. Zinchenko, Yu.F. Lonin, A.G. Ponomarev, and S.I. Fedotov. Problems of Atomic Science and Technology. Series: Plasma electronics and new acceleration methods., (4(86)), 7–9 (2013).

[11] A.B. Batrakov, I.N. Onishchenko, S.I. Fedotov, et al. . Problems of Atomic Science and Technology. (1(155)), 61-64 (2025). https://doi.org/10.46813/2025-155-061

[12] R.J. Thome, and J.M. Tarrh, MHD and Fusion Magnets, (JWS, NY, 1982). pp. 247-248.

[13] https://www.radpro-int.com/tld-1/rados-tld-reader/

[14] E. Villegas, and F. Somarriba, . International Conference on Radiation Safety: Improving Radiation Protection in Practice. Extended Abstracts, IAEA-CN-279 305–306 (2021).
\end{flushleft}

\begin{center}
\fontsize{10}{14}\selectfont
\selectlanguage{ukrainian}
\noindent \textbf{ВИЗНАЧЕННЯ ЗАГАЛЬНОЇ ДОЗИ ГАЛЬМІВНОГО РЕНТГЕНІВСЬКОГО ВИПРОМІНЮВАННЯ НА СИЛЬНОСТРУМОВОМУ ІМПУЛЬСНОМУ ВИПРОМІНЮВАЛЬНО-ПУЧКОВОМУ КОМПЛЕКСІ «TEMP-B»}

\fontsize{10}{14}\selectfont
\selectlanguage{ukrainian}
\noindent \textbf{А.Б. Батраков, С.І. Федотов, І.М. Оніщенко, Є.Г. Глушко, А.М. Горбань, О.Л. Рак,}

\noindent \textbf{О.В. Невара, Ю.М. Волков}

\selectlanguage{ukrainian}
\noindent {\emph{Національний науковий центр «Харківський фізико-технічний інститут», Харків, Україна}}
\end{center}

\selectlanguage{ukrainian}
 \noindent Представлено експериментальне вимірювання повної дози потужного джерела гальмівного рентгенівського випромінювання, базованого на імпульсному прискорювачі електронів прямої дії з параметрами сильнострумового релятивістського трубчастого електронного пучка: енергія 600 кеВ, струм 13,5 кА і тривалість імпульсу 1,0 мкс. Транспортування електронного пучка, отримуваного в магнітоізольованому діоді, до молібденового конвертера на відстань 55 см здійснювалось імпульсним магнітним полем соленоїда. До нього примикає додаткова котушка, з'єднана послідовно з ним, для забезпечення магнітного поля в області конвертера та уникнення втрат пучка. Описана методика визначення повної дози гальмівного рентгенівського випромінювання. Для вимірювання дози гальмівного рентгенівського випромінювання використовувалися полікристалічні детектори, які розташовувалися на відстані 80 мм за конвертором. Детектори розміщувалися в полярній площині з кроком $20^\circ$. Виміри розподілу доз за полярним кутом показали симетрію випромінювання за полярним та азимутальним кутами відносно оси. Беручи до уваги, що пучок випромінює тільки передню напівсферу, то разом з виміряною симетрією можливо інтегрування по всій сфері звести до інтегрування лише по 1/8 площині сфери, зменшивши кількість датчиків. Експериментально отримана величина повної дози гальмівного рентгенівського випромінювання 388,5 Гр за один імпульс струму прискорювача, яка зосереджена в куті $120^\circ$ в напрямку руху пучка.

\begin{flushleft}
\fontsize{10}{14}\selectfont
\selectlanguage{ukrainian}
\noindent \textbf{Ключові слова:} \emph{ сильнострумовий імпульсний електронний прискорювач прямої дії; релятивістський електронний пучок; конвертор; гальмівне рентгенівське випромінювання; термолюмінесцентний детектор; інтенсивність; доза}
 \end{flushleft}

\end{document}